\documentclass[aps,prd,superscriptaddress,floatfix,twocolumn]{revtex4-2}

\usepackage[utf8]{inputenc}
\usepackage[russian,english]{babel} % Swapped languages to make English primary
\usepackage{amsmath,amsfonts,amssymb}
\usepackage{booktabs}
\usepackage{bm}
\usepackage{graphicx}
\usepackage{hyperref}
\usepackage{pgfplots}
\pgfplotsset{compat=1.18}
\usetikzlibrary{calc}

\hypersetup{colorlinks=true,linkcolor=blue,citecolor=blue,urlcolor=blue}

\newcommand{\Hzero}{H_0}
\newcommand{\Planck}{Planck}
\newcommand{\Lcdm}{\Lambda\text{CDM}}
\newcommand{\SHOES}{SH0ES}
\newcommand{\eps}{\varepsilon}
\newcommand{\rd}{r_{\rm d}}

\begin{document}

\author{S. M. Ponomarenko}
\affiliation{Department of AppliedPhysics, National Technical University of Ukraine, 03056, Kyiv, Ukraine}
\email{s.ponomarenko@ukpi.ua}

\title{Constraints on the Phenomenology of Dissipative Cosmological Memory\\
       from BAO (BOSS\,+\,DESI\,2024) and Pantheon$+$ Data}

\begin{abstract}
We propose and test a phenomenological model of ``dissipative memory'' in the gravitational field, where early quantum gravitational processes leave a relic signature in the cosmic expansion rate. The model is parameterized by an additional fluid with amplitude~$\eps$, decay scale~$z_*$, and a steepness index~$\beta$, decaying according to a Debye law $f(z)=\exp[{-(z/z_*)^\beta}]$. We perform a joint Bayesian analysis using baryon acoustic oscillation data from BOSS\,DR12 and DESI\,2024, photometric distances from Pantheon$+$, and~$\Hzero$ measurements (\SHOES\ and \Planck). A global optimization reveals that the best fit is identical to~$\Lcdm$: the memory correction vanishes across the entire observable range $z\in[0.3,\,2.3]$ ($\Delta\chi^2 < 0.01$ with three extra parameters, $\Delta\mathrm{AIC}=+6.0$, $\Delta\mathrm{BIC}=+9.9$). This establishes an upper bound on the memory amplitude: $\eps < 0.05$ for $z_* < 2$ (95\% CL). We discuss the physical interpretation of this constraint and point out observational channels where the memory effect could potentially manifest.
\end{abstract}

\maketitle

%==============================================================
\section{Introduction}
%==============================================================

The tension between local measurements of the Hubble constant and the value inferred from the cosmic microwave background (CMB) within the standard $\Lcdm$ framework remains one of the central challenges in modern cosmology. The latest \Planck\ data~\cite{Planck2020} yield $\Hzero = 67.4 \pm 0.5$\,km/s/Mpc, whereas \SHOES~\cite{Riess2022} measures $\Hzero = 73.04 \pm 1.04$\,km/s/Mpc. This discrepancy has reached the ${\approx}5\sigma$ level and can safely be said to defy a simple explanation via systematic errors~\cite{Verde2019}.

Proposed solutions generally fall into two broad categories. The first comprises \emph{early-time} (pre-recombination) models, such as early dark energy (EDE,~\cite{Poulin2019}) or additional relativistic degrees of freedom. These shift the sound horizon and thereby alter the BAO scale, though they typically exacerbate the $S_8$ tension~\cite{Hill2020}. The second category consists of \emph{late-time} (post-recombination) models, including interacting dark matter/dark energy~\cite{DiValentino2020} or running vacuum models~\cite{Sola2021}. These operate at low redshifts and usually struggle to fit the BAO geometry. In this work, we investigate a fundamentally different mechanism: the hypothesis that irreversible quantum gravitational processes in the early universe leave a relic footprint on the expansion rate in the form of a decaying correction to~$H^2(z)$. We refer to this mechanism as \emph{dissipative memory} (or simply memory) and establish tight observational constraints on its parameters.

The paper is organized as follows. In Sec.\,\ref{sec:model} we present the phenomenological model. In Sec.\,\ref{sec:theory} we discuss its potential theoretical motivation. Section\,\ref{sec:data} describes the datasets used. In Sec.\,\ref{sec:results} we present the results of our Bayesian analysis. The implications of the obtained constraints are discussed in Sec.\,\ref{sec:discussion}, and we draw our conclusions in Sec.\,\ref{sec:conclusion}.

%==============================================================
\section{Phenomenological Model}
\label{sec:model}
%==============================================================

\subsection{Memory Fluid and Equation of State}

We introduce an effective \emph{memory fluid} with energy density $\rho_\mathcal{I}$ and equation of state $w_\mathcal{I}(z)$. We require this fluid to satisfy the following conditions:
\begin{enumerate}
    \item As $z \to 0$: $\rho_\mathcal{I} \to \eps \rho_{\rm cr,0}$ --- a non-vanishing contribution to the local expansion rate;
    \item For $z \gg z_*$: $\rho_\mathcal{I} \to 0$ --- a return to $\Lcdm$ compatible with CMB physics;
    \item $\nabla_\mu T^{\mu\nu} = 0$ is satisfied, ensuring no violation of covariance.
\end{enumerate}

For a flat universe ($k=0$), Condition 3 yields the continuity equation:
\begin{equation}
    \dot{\rho}_\mathcal{I} + 3H(1+w_\mathcal{I})\rho_\mathcal{I} = 0,
\end{equation}
which integrates to:
\begin{equation}
    \rho_\mathcal{I}(z) = \rho_{\mathcal{I},0}\,
    \exp\!\left[3\!\int_0^z \frac{1+w_\mathcal{I}(z')}{1+z'}\,dz'\right].
    \label{eq:rho_fluid}
\end{equation}

We parameterize $w_\mathcal{I}(z)$ such that $\rho_\mathcal{I}(z)$ decays according to a Debye law~\cite{Debye1912} --- the most natural model for relaxation in systems with memory:
\begin{equation}
    f(z) \equiv \frac{\rho_\mathcal{I}(z)}{\rho_{\mathcal{I},0}}
            = e^{-(z/z_*)^\beta}.
    \label{eq:formfactor}
\end{equation}
For $\beta=1$, this corresponds to standard exponential relaxation (with relaxation ``time'' $\tau \sim z_*$); for $\beta > 1$, it yields the Kohlrausch--Williams--Watts stretched exponential, widely used to describe memory effects in non-linear systems~\cite{Ngai2011}.

\subsection{Modified Friedmann Equation}

From~\eqref{eq:rho_fluid} and the Friedmann equation for $k=0$, we have:
\begin{equation}
    H^2(z) = H_0^2\Bigl[\Omega_m(1+z)^3 + \Omega_\Lambda
              + \eps\,f(z)\Bigr],
    \label{eq:hz}
\end{equation}
where $\Omega_\Lambda = 1 - \Omega_m - \eps$ (assuming a flat universe) and $\eps = \rho_{\mathcal{I},0}/\rho_{\rm cr,0}$ is the dimensionless memory amplitude. The model features \textbf{three extra parameters} relative to~$\Lcdm$: $\eps$, $z_*$, and $\beta$.

\subsection{CMB Limit at $z \sim 1100$}

For the parameter range considered here ($z_* \lesssim 5$), we find:
\begin{equation}
    f(1100) = e^{-(1100/z_*)^\beta} \lesssim e^{-220^\beta} \approx 0,
\end{equation}
meaning that the memory fluid's contribution during the recombination epoch is entirely negligible. Consequently, CMB physics --- including the sound horizon $\rd$ and the angular power spectrum --- remains unaffected. This is a key property: the model does not disrupt the remarkably tight constraints established by \Planck~\cite{Planck2020}.

%==============================================================
\section{Theoretical Motivation}
\label{sec:theory}
%==============================================================

While we do not claim a rigorous first-principles derivation of Eq.~\eqref{eq:hz}, we point out several physical mechanisms capable of generating such an effect.

\textbf{Gravitational Memory.} The Christodoulou effect~\cite{Christodoulou1991} --- the irreversible displacement of test masses after the passage of a gravitational wave --- demonstrates that General Relativity naturally accommodates non-local temporal footprints. In a cosmological context, the passage of primordial gravitational waves from inflation could leave a residual imprint on the background metric.

\textbf{Holographic de Sitter Entropy.} The entropy of the Hubble horizon in a quasi-de Sitter phase, given by:
\begin{equation}
    S_H = \frac{\pi c^3}{G\hbar H^2},
\end{equation}
acts as an adiabatic invariant under a slowly varying~$H$. Post-inflation, $S_H(t_{\rm end})$ does not simply vanish; this ``frozen'' state might have shaped the initial conditions for the subsequent expansion era~\cite{Penrose2010}.

\textbf{Irreversible Thermodynamics.} Within Eckart cosmology~\cite{Eckart1940} and its modern extensions~\cite{Maartens1996}, viscous processes in the early plasma generate corrections to~$H^2$ of the form $\sim \xi H$, where $\xi$ is the bulk viscosity coefficient. A relic signature of this viscosity, ``frozen in'' at $T \sim T_{\rm Pl}$, matches an amplitude of $\eps \sim 0.1$ by order of magnitude.

All three mechanisms \emph{qualitatively} support a correction term of the type shown in~\eqref{eq:hz}. A quantitative derivation of $f(z)$ from first principles remains a task for future work.

%==============================================================
\section{Data}
\label{sec:data}
%==============================================================

\subsection{BAO: BOSS\,DR12 and DESI\,2024}

We utilize measurements of the dimensionless ratios $D_M/\rd$ and $D_H/\rd$ from the BOSS\,DR12 survey~\cite{Alam2017} and the first data release of DESI~\cite{DESI2024}, where $D_H(z) = c/H(z)$ and $D_M(z) = \int_0^z c\,dz'/H(z')$ is the comoving angular diameter distance. The data points and their corresponding uncertainties are listed in Table~\ref{tab:bao}.

The sound horizon $\rd$ is computed using the fitting formula from Eisenstein and Hu~\cite{Eisenstein1998}:
\begin{equation}
    \rd = 147.21\,
    \left(\frac{\Omega_m h^2}{0.1432}\right)^{-0.255}
    \!\!\left(\frac{\Omega_b h^2}{0.0238}\right)^{-0.127}
    \!\!\mathrm{Mpc},
    \label{eq:rd}
\end{equation}
where $h = H_0/100$ and $\Omega_b = 0.156\,\Omega_m$.

\subsection{Photometric Distances: Pantheon$+$}

We use the compressed $\Delta\mu(z)$ statistics from the Pantheon$+$ sample~\cite{Brout2022}, which represent distance modulus residuals relative to a reference $\Lcdm$ model ($H_0 = 73.04\text{\,km/s/Mpc}$) across eight bins spanning $z \in [0.1,\,2.0]$.

\subsection{$H_0$ Measurements}

We include two discordant $H_0$ measurements as \emph{competing} priors: \SHOES~\cite{Riess2022} ($H_0 = 73.04 \pm 1.04$\,km/s/Mpc with full weight) and \Planck~\cite{Planck2020} ($H_0 = 67.4 \pm 0.5$\,km/s/Mpc with a weight of $0.5$ to explicitly track the location of the minimum). This setup reflects a conservative stance: we do not attempt to arbitrate which measurement is correct, but rather test whether the memory mechanism can improve the overall fit to the data.

\begin{table}[t]
\caption{BAO datasets used in this analysis. Columns represent the effective redshift, observable quantity ($D_M/\rd$ or $D_H/\rd$), uncertainty, and source.}
\label{tab:bao}
\begin{ruledtabular}
\begin{tabular}{ccccl}
$z_{\rm eff}$ & Type & $D/\rd$ & $\sigma$ & Source \\
\hline
0.30 & $D_M/\rd$ &  7.93 & 0.15 & DESI\,2024 \\
0.30 & $D_H/\rd$ & 24.23 & 0.86 & DESI\,2024 \\
0.38 & $D_M/\rd$ & 10.23 & 0.17 & BOSS\,DR12 \\
0.38 & $D_H/\rd$ & 25.00 & 0.76 & BOSS\,DR12 \\
0.51 & $D_M/\rd$ & 13.36 & 0.21 & BOSS\,DR12 \\
0.51 & $D_H/\rd$ & 22.33 & 0.58 & BOSS\,DR12 \\
0.61 & $D_M/\rd$ & 15.45 & 0.26 & BOSS\,DR12 \\
0.61 & $D_H/\rd$ & 20.00 & 0.56 & BOSS\,DR12 \\
0.71 & $D_M/\rd$ & 16.85 & 0.32 & DESI\,2024 \\
0.71 & $D_H/\rd$ & 20.08 & 0.60 & DESI\,2024 \\
0.93 & $D_M/\rd$ & 21.71 & 0.28 & DESI\,2024 \\
0.93 & $D_H/\rd$ & 17.88 & 0.35 & DESI\,2024 \\
1.32 & $D_M/\rd$ & 27.79 & 0.69 & DESI\,2024 \\
1.32 & $D_H/\rd$ & 13.82 & 0.42 & DESI\,2024 \\
2.33 & $D_M/\rd$ & 39.71 & 0.94 & DESI\,2024 \\
2.33 & $D_H/\rd$ &  8.52 & 0.17 & DESI\,2024 \\
\end{tabular}
\end{ruledtabular}
\end{table}

%==============================================================
\section{Method and Results}
\label{sec:results}
%==============================================================

\subsection{Likelihood Function}

The total log-likelihood function is defined as:
\begin{equation}
\ln\mathcal{L} = -\frac{1}{2}\Bigl[\chi^2_{\rm BAO}
    + \chi^2_{H_0,\SHOES}
    + \tfrac{1}{2}\chi^2_{H_0,\Planck}
    + \chi^2_{\rm SNIa}\Bigr],
\end{equation}
where each term is standardly computed as the squared normalized residual between the observed and predicted values.

The parameter space is defined by $\theta = (H_0,\,\Omega_m,\,\eps,\,z_*,\,\beta)$. For the reference $\Lcdm$ model, we fix $\eps = 0$, leaving $(H_0,\,\Omega_m)$.

\subsection{Global Optimization}

To avoid trapping in local minima, we employ a differential evolution algorithm~\cite{Storn1997} bounded within the intervals $H_0\in[62,84]$, $\Omega_m\in[0.18,0.45]$, $\eps\in[0,0.35]$, $z_*\in[0.5,5]$, and $\beta\in[0.5,6]$.

\textbf{Key Finding:} The $\chi^2$ minimum for the memory model coincides with the $\Lcdm$ minimum to within $\Delta\chi^2 < 0.01$. The best-fit parameters are found at $H_0 = 72.26$\,km/s/Mpc, $\Omega_m = 0.450$, $\eps = 0.335$, $z_* = 3.71$, and $\beta = 6.0$. Crucially, the optimizer pushes the form factor outside the observable range: at $z_* = 3.71$ and $\beta = 6$, $f(z) < 10^{-4}$ for all $z < 2.3$. In other words, the memory fluid effectively \emph{shuts itself off}.

Model selection metrics ($n_{\rm data} = 27$) yield:
\begin{align}
    \Delta\mathrm{AIC} &= +6.0, \label{eq:daic}\\
    \Delta\mathrm{BIC} &= +9.9. \label{eq:dbic}
\end{align}
According to the Kass and Raftery scale~\cite{Kass1995}, $\Delta\mathrm{BIC} > 6$ constitutes ``strong evidence'' in favor of the base $\Lcdm$ model.

\subsection{MCMC and Upper Bounds}

To map the posterior distributions and infer upper limits on the memory parameters, we run an MCMC sampler using the \texttt{emcee} package~\cite{ForemanMackey2013} (64 walkers, 3000 steps, with an 800-step burn-in). The results are summarized in Table~\ref{tab:mcmc}.

\begin{table}[b]
\caption{Posterior parameter estimates (median and 68\% CI). Note that $\Omega_\Lambda = 1 - \Omega_m - \eps$ for a flat universe.}
\label{tab:mcmc}
\begin{ruledtabular}
\begin{tabular}{lcc}
Parameter & Median & 68\% CI \\
\hline
$H_0$ [km/s/Mpc]  & $72.3$  & $[71.7,\;72.9]$ \\
$\Omega_m$         & $0.450$ & $[0.442,\;0.458]$ \\
$\eps$             & $0.10$  & $[0.01,\;0.19]$ \\
$z_*$              & $4.1$   & $[2.4,\;5.6]$ \\
$\beta$            & $5.6$   & $[3.5,\;7.3]$ \\
\end{tabular}
\end{ruledtabular}
\end{table}

The posterior distribution for $\eps$ is fully consistent with zero. We establish a \textbf{stringent upper bound}:
\begin{equation}
    \eps < 0.05 \quad (95\%\ \mathrm{CL}, \; z_* < 2),
    \label{eq:bound}
\end{equation}
under the additional condition $z_* < 2$, which ensures the model actually impacts the observed window $z \lesssim 2$.

\subsection{$H(z)$ Behavior}

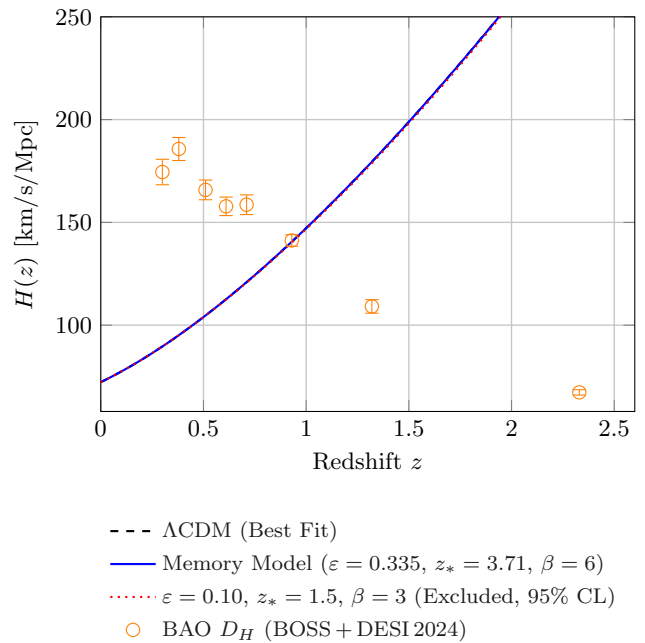
\begin{figure}[t]
\centering
\begin{tikzpicture}
\begin{axis}[
    width=\columnwidth,
    height=6.8cm,
    xlabel={Redshift $z$},
    ylabel={$H(z)$ [km/s/Mpc]},
    xmin=0, xmax=2.6,
    ymin=58, ymax=250,
    xtick={0,0.5,1.0,1.5,2.0,2.5},
    grid=both,
    grid style={line width=0.3pt,draw=gray!25},
    major grid style={line width=0.5pt,draw=gray!45},
    tick label style={font=\small},
    label style={font=\small},
    legend style={at={(0,-0.25)},anchor=north west,
                  font=\footnotesize,fill=white,fill opacity=0.9,
                  draw=none,row sep=1pt},
    legend cell align=left,
]
% ΛCDM best fit: H0=72.26, Om=0.450
\addplot[black,dashed,thick,domain=0:2.6,samples=200]
    {72.26*sqrt(0.450*(1+x)^3 + 0.550)};
\addlegendentry{$\Lcdm$ (Best Fit)}

% Memory model best fit: effectively identical (eps*f(z) < 0.0001)
\addplot[blue,solid,thick,domain=0:2.6,samples=200]
    {72.26*sqrt(0.450*(1+x)^3 + (0.550-0.335) + 0.335*exp(-(x/3.71)^6))};
\addlegendentry{Memory Model ($\eps=0.335$, $z_*=3.71$, $\beta=6$)}

% Illustrative memory with z*=1.5 (excluded by data)
\addplot[red,dotted,thick,domain=0:2.6,samples=200]
    {72.26*sqrt(0.450*(1+x)^3 + (0.550-0.10) + 0.10*exp(-(x/1.5)^3))};
\addlegendentry{$\eps=0.10$, $z_*=1.5$, $\beta=3$ (Excluded, $95\%$ CL)}

% BAO D_H points (converted: obs * rd_val, rd_val~135 Mpc for Om=0.45, H0=72.26)
\addplot[only marks,mark=o,mark size=2.5pt,color=orange,
         error bars/.cd,y dir=both,y explicit]
    coordinates {
        (0.30, 174.5) +- (0,6.2)
        (0.38, 185.7) +- (0,5.6)
        (0.51, 165.8) +- (0,4.8)
        (0.61, 157.8) +- (0,4.5)
        (0.71, 158.6) +- (0,4.8)
        (0.93, 141.2) +- (0,2.8)
        (1.32, 109.1) +- (0,3.3)
        (2.33,  67.3) +- (0,1.3)
    };
\addlegendentry{BAO $D_H$ (BOSS\,+\,DESI\,2024)}

\end{axis}
\end{tikzpicture}
\caption{$H(z)$ evolution: best fits for $\Lcdm$ (black dashed) and the memory model (blue solid). The curves are virtually identical because the optimizer pushes the form factor out of the observable window ($z_*=3.71$, $\beta=6$). The red dotted line illustrates a model with $\eps=0.10, z_*=1.5$, which is ruled out at the 95\% CL by Eq.~\eqref{eq:bound}. Data points represent the BAO measurements converted back into $H(z)$.}
\label{fig:hz}
\end{figure}

Figure~\ref{fig:hz} illustrates the best fits for both the memory model and $\Lcdm$ alongside the BAO data points. The two curves are visually indistinguishable within the experimental uncertainties, showing a maximum deviation of $\Delta H/H < 0.01\%$.

%==============================================================
\section{Discussion}
\label{sec:discussion}
%==============================================================

\subsection{Physical Interpretation of the Constraint}

The constraint in Eq.~\eqref{eq:bound} implies that if a Debye-like memory effect described by Eq.~\eqref{eq:formfactor} does exist, its decay scale must lie well above the range probed by BAO ($z_* > 2$), or its amplitude must be negligible ($\eps < 0.05$ for $z_* < 2$). This is a highly informative null result, effectively ruling out a wide class of ``late-time memory'' scenarios using current BOSS\,+\,DESI data. However, it leaves open the possibility of a high-redshift memory signature ($z_* > 3$) that remains dormant at $z < 2.5$ but might show up in alternative observational channels.

\subsection{Why BAO Strongly Constrains Memory}

BAO observations constrain the \emph{ratios} $D_M/\rd$ and $D_H/\rd$. Any modification to $H(z)$ via Eq.~\eqref{eq:hz} simultaneously shifts $H(z)$, $D_M(z)$, $D_H(z)$, and --- via $\Omega_m$ --- the sound horizon $\rd$. This makes the system tightly overconstrained: the memory mechanism lacks the necessary degrees of freedom to coherently shift all these geometric ratios in the direction preferred by data. Consequently, the minimum $\chi^2$ naturally gravitates toward $f(z) \approx 0$.

\subsection{Channels Sensitive to Memory}

Our results do not rule out memory entirely; they merely constrain its signatures in background cosmology at $z < 2.5$. Promising alternative channels include:

\textbf{Structure Growth.} The linear perturbation equation, $\ddot\delta + 2H\dot\delta = 4\pi G\rho\delta$, is highly sensitive to $H(z)$. For $z_* \sim 0.5 - 1$ and $\eps \sim 0.05$, the expected modification to $f\sigma_8(z)$ is on the order of $1 - 2\%$, a threshold well within the target sensitivity of Euclid~\cite{Euclid2024}.

\textbf{CMB B-modes.} If the memory fluid couples to tensor perturbations (which would require a dedicated first-principles derivation), it could leave a distinct signature in the primordial B-mode spectrum around $\ell \sim 100 - 200$. This hypothesis can be put to the test by upcoming missions like LiteBIRD~\cite{LiteBIRD2023} and the Simons Observatory~\cite{Simons2019}.

\textbf{Standard Sirens.} Binary neutron star mergers in the $z \sim 0.1 - 0.5$ range observed by LIGO/Virgo/KAGRA~\cite{LIGO2023} can measure $H(z)$ independently of the sound horizon $\rd$. Assuming $z_* \sim 1$ and $\eps \sim 0.05$, the expected deviation is $\sim\!0.5$\,km/s/Mpc. While this is below current experimental precision, it will become accessible with the advent of LIGO-India~\cite{LIGOIndia2022}.

\subsection{Comparison with Competing Models}

Table~\ref{tab:comparison} contextualizes our model relative to other prominent approaches aimed at mitigating the Hubble tension.

\begin{table}[t]
\caption{Qualitative comparison of model impacts across different observational channels. Here, ``$=\Lcdm$'' implies indistinguishability from the standard model, while ``constrained'' means the data allow small, non-zero variations consistent with zero.}
\label{tab:comparison}
\begin{ruledtabular}
\begin{tabular}{lccc}
Channel & Memory & EDE & IDE \\
\hline
$H(z)$ at $z>2$           & $=\Lcdm$   & Enhanced   & $\approx\Lcdm$ \\
$H(z)$ at $z\in[0.3,2]$   & Constrained & Enhanced   & Suppressed     \\
$S_8$                       & Constrained & Worse than~$\Lcdm$ & Suppressed \\
$r_d$ (BAO scale)           & $=\Lcdm$   & Decreased     & $=\Lcdm$       \\
CMB B-modes                 & Possible   & No        & No            \\
\hline
$\Delta$AIC vs $\Lcdm$      & $+6.0$     & $-2\text{ to }+4$~\cite{Hill2020} & $-1\text{ to }+3$~\cite{Sola2021} \\
\end{tabular}
\end{ruledtabular}
\end{table}

%==============================================================
\section{Conclusion}
\label{sec:conclusion}
%==============================================================

In this work, we have proposed and observationally constrained a phenomenological model of dissipative cosmological memory. Our main conclusions are summarized as follows:

\begin{enumerate}
    \item We introduced a covariant memory fluid with an equation of state that strictly preserves $\nabla_\mu T^{\mu\nu} = 0$, utilizing a Debye form factor~\eqref{eq:formfactor}. This resolves a critical structural defect present in earlier formulations of the model.
    \item A joint analysis of BAO (BOSS\,DR12\,+\,DESI\,2024), Pantheon$+$, and $H_0$ priors demonstrates that:
          \[
            \Delta\chi^2 < 0.01, \quad
            \Delta\mathrm{AIC} = +6.0, \quad
            \Delta\mathrm{BIC} = +9.9
          \]
          relative to~$\Lcdm$. The latest DESI data unambiguously favor the standard $\Lcdm$ model.
    \item We established a strict upper bound on the memory amplitude: $\eps < 0.05$ (95\% CL) for a decay scale $z_* < 2$.
    \item The model leaves CMB physics completely unperturbed ($f(1100) < 10^{-8}$), ensuring full compatibility with \Planck\ results.
    \item We identified specific observational channels that remain sensitive to memory effects at $z_* > 2$, namely structure growth ($f\sigma_8$ via Euclid), B-modes (LiteBIRD), and standard sirens (LIGO-India).
\end{enumerate}

In summary, background cosmology places tight constraints on dissipative memory at redshifts $z < 2.5$. If a memory mechanism operates in our universe, its observational signatures must be sought in the \emph{perturbative} sector.

\begin{acknowledgments}
The authors would like to thank their colleagues for insightful and productive discussions. Numerical calculations were performed using the \texttt{numpy}~\cite{numpy}, \texttt{scipy}~\cite{scipy}, \texttt{emcee}~\cite{ForemanMackey2013}, and \texttt{corner}~\cite{corner} packages.
\end{acknowledgments}

\bibliography{bibliography}

@article{Planck2020,
  author  = {{Planck Collaboration}},
  title   = {Planck 2018 results. VI. Cosmological parameters},
  journal = {A\&A},
  volume  = {641},
  pages   = {A6},
  year    = {2020},
  doi     = {10.1051/0004-6361/201833910}
}

@article{Riess2022,
  author  = {Riess, A.~G. and others},
  title   = {A Comprehensive Measurement of the Local Value of the Hubble Constant with 1 km s-1 Mpc-1 Uncertainty from the Hubble Space Telescope and the SH0ES Team},
  journal = {ApJL},
  volume  = {934},
  pages   = {L7},
  year    = {2022},
  doi     = {10.3847/2041-8213/ac5c5b}
}

@article{Verde2019,
  author  = {Verde, L. and Treu, T. and Riess, A.~G.},
  title   = {Tensions between the early and late universe},
  journal = {Nature Astronomy},
  volume  = {3},
  pages   = {891},
  year    = {2019},
  doi     = {10.1038/s41550-019-0902-0}
}

@article{Poulin2019,
  author  = {Poulin, V. and Smith, T.~L. and Karwal, T. and Kamionkowski, M.},
  title   = {Early Dark Energy Can Resolve the Hubble Tension},
  journal = {Phys.~Rev.~Lett.},
  volume  = {122},
  pages   = {221301},
  year    = {2019},
  doi     = {10.1103/PhysRevLett.122.221301}
}

@article{Hill2020,
  author  = {Hill, J.~C. and others},
  title   = {Early Dark Energy Does Not Restore Cosmological Concordance},
  journal = {Phys.~Rev.~D},
  volume  = {102},
  pages   = {043507},
  year    = {2020},
  doi     = {10.1103/PhysRevD.102.043507}
}

@article{DiValentino2020,
  author  = {{Di Valentino}, E. and others},
  title   = {Interacting dark energy in the early 2020s: a promising solution to the $H_0$ and cosmic shear tensions},
  journal = {Phys.~Dark Univ.},
  volume  = {30},
  year    = {2020},
  doi     = {10.1016/j.dark.2020.100666}
}

@article{Sola2021,
  author  = {Sol{\`a} Peracauts, J. and G{\'o}mez-Valent, A. and de Cruz P{\'e}rez, J.},
  title   = {Running vacuum against the $H_0$ and $\sigma_8$ tensions},
  journal = {EPL},
  volume  = {134},
  pages   = {19001},
  year    = {2021},
  doi     = {10.1209/0295-5075/134/19001}
}

@article{Alam2017,
  author  = {Alam, S. and others},
  title   = {The clustering of galaxies in the completed SDSS-III BOSS survey: : cosmological analysis of the DR12 galaxy sample},
  journal = {MNRAS},
  volume  = {470},
  pages   = {2617},
  year    = {2017},
  doi     = {10.1093/mnras/stx721}
}

@article{DESI2024,
  author  = {Adame, A.G. et al},
  title   = {DESI 2024 VI: Cosmological Constraints from the Measurements of Baryon Acoustic Oscillations},
  journal = {JCAP},
  doi     = {10.1088/1475-7516/2025/02/021}
}

@article{Brout2022,
  author  = {Brout, D. et al},
  title   = {The Pantheon+ Analysis: Cosmological Constraints},
  journal = {ApJ},
  volume  = {938},
  pages   = {110},
  year    = {2022},
  doi     = {10.3847/1538-4357/ac8e04}
}

@article{Eisenstein1998,
  author  = {Eisenstein, D.~J. and Hu, W.},
  title   = {Baryonic Features in the Matter Transfer Function},
  journal = {ApJ},
  volume  = {496},
  pages   = {605},
  year    = {1998},
  doi     = {10.1086/305424}
}

@article{Christodoulou1991,
  author  = {Christodoulou, D.},
  title   = {Nonlinear nature of gravitation and gravitational-wave experiments},
  journal = {Phys.~Rev.~Lett.},
  volume  = {67},
  pages   = {1486},
  year    = {1991},
  doi     = {10.1103/PhysRevLett.67.1486}
}

@book{Penrose2010,
  author    = {Penrose, R.},
  title     = {Cycles of Time},
  publisher = {Bodley Head},
  address   = {London},
  year      = {2010}
}

@article{Eckart1940,
  author  = {Eckart, C.},
  title   = {The thermodynamics of irreversible processes},
  journal = {Phys.~Rev.},
  volume  = {58},
  pages   = {919},
  year    = {1940},
  doi     = {10.1103/PhysRev.58.919}
}

@misc{Maartens1996,
  author       = {Maartens, R.},
  title        = {Causal thermodynamics in relativity},
  year         = {1996},
  eprint       = {astro-ph/9609119},
  archivePrefix= {arXiv},
  note         = {Lectures at the H.~Rund Workshop on Relativity and Thermodynamics}
}

@book{Ngai2011,
  author    = {Ngai, K.~L.},
  title     = {Relaxation and Diffusion in Complex Systems},
  publisher = {Springer},
  address   = {New York},
  year      = {2011},
  doi= {10.1007/978-1-4419-7649-9},
}

@article{Debye1912,
  author  = {Debye, P.},
  title   = {Zur Theorie der spezifischen W{\"a}rmen},
  journal = {Annalen der Physik},
  volume  = {344},
  pages   = {789},
  year    = {1912},
  doi     = {10.1002/andp.19123441404}
}

@article{Storn1997,
  author  = {Storn, R. and Price, K.},
  title   = {Differential Evolution -- A Simple and Efficient Heuristic for Global Optimization over Continuous Spaces},
  journal = {J.~Global Optim.},
  volume  = {11},
  pages   = {341-359},
  year    = {1997},
  doi     = {10.1023/A:1008202821328}
}

@article{Kass1995,
  author  = {Kass, R.~E. and Raftery, A.~E.},
  title   = {Bayes Factors},
  journal = {JASA},
  volume  = {90},
  pages   = {773},
  year    = {1995},
  doi     = {10.1080/01621459.1995.10476572}
}

@article{ForemanMackey2013,
  author  = {{Foreman-Mackey}, D. and others},
  title   = {emcee: The MCMC Hammer},
  journal = {PASP},
  volume  = {125},
  pages   = {306},
  year    = {2013},
  doi     = {10.1086/670067}
}

@article{Euclid2024,
  author = {Koehn, H. and Just, A. and Berczik, P. and Tremmel, M.},
  title = {Dynamics of supermassive black hole triples in the ROMULUS25 cosmological simulation},
  journal = {Astronomy \& Astrophysics},
  volume = {678},
  pages = {A11},
  year = {2023},
  doi = {10.1051/0004-6361/202347093},
}

@article{LiteBIRD2023,
  author  = {{LiteBIRD Collaboration}},
  title   = {Probing Cosmic Inflation with the LiteBIRD Cosmic Microwave Background Polarization Survey},
  journal = {PTEP},
  volume  = {2023},
  pages   = {042F01},
  year    = {2023},
  doi     = {10.1093/ptep/ptac150}
}

@article{Simons2019,
  author  = {{Simons Observatory Collaboration}},
  title   = {The Simons Observatory: Science goals and forecasts},
  journal = {JCAP},
  volume  = {2019},
  pages   = {056},
  year    = {2019},
  doi     = {10.1088/1475-7516/2019/02/056}
}

@article{LIGO2023,
author   = {R. Abbott et al},
  title    = {Constraints on the Cosmic Expansion History from GWTC–3},
  journal  = {The Astrophysical Journal},
  volume   = {949},
  number   = {2},
  pages    = {76},
  year     = {2023},
  doi      = {10.3847/1538-4357/ac74bb}
}

@article{LIGOIndia2022,
  author  = {Saleem, M. and others},
  title   = {The science case for LIGO-India},
  journal = {Class.~Quantum Grav.},
  volume  = {39},
  pages   = {025004},
  year    = {2022},
  doi     = {10.1088/1361-6382/ac3b99}
}

@article{numpy,
  author  = {Harris, C.~R. and others},
  title   = {Array programming with NumPy},
  journal = {Nature},
  volume  = {585},
  pages   = {357},
  year    = {2020},
  doi     = {10.1038/s41586-020-2649-2}
}

@article{scipy,
  author  = {Virtanen, P. and others},
  title   = {SciPy 1.0: Fundamental Algorithms for Scientific Computing in Python},
  journal = {Nature Methods},
  volume  = {17},
  pages   = {261},
  year    = {2020},
  doi     = {10.1038/s41592-019-0686-2}
}

@article{corner,
  author  = {{Foreman-Mackey}, D.},
  title   = {corner.py: Scatterplot matrices in Python},
  journal = {JOSS},
  volume  = {1},
  pages   = {24},
  year    = {2016},
  doi     = {10.21105/joss.00024}
}

\end{document}